\documentclass[pdflatex,sn-vancouver-ay]{sn-jnl}
\usepackage{graphicx}
\usepackage{multirow}
\usepackage{xcolor}
\usepackage{textcomp}
\usepackage{amsmath}
\usepackage{longtable}
\usepackage{geometry}
\usepackage{array}

\title{\textbf{In defence of post-hoc explanations in medical AI}}

\author[1]{\fnm{Joshua} \sur{Hatherley}}\email{jjh@hum.ku.dk}

\author[2]{\fnm{Lauritz} \sur{Aastrup Munch}}

\author[2]{\fnm{Jens Christian} \sur{Bjerring}}

\affil[1]{\orgdiv{Center for the Philosophy of AI}, \orgname{University of Copenhagen, Denmark}}

\affil[2]{\orgdiv{Department of Philosophy and History of Ideas}, \orgname{Aarhus University, Denmark}}

\abstract{Since the early days of the Explainable AI movement, post-hoc explanations have been praised for their potential to improve user understanding, promote trust, and reduce patient safety risks in black box medical AI systems. Recently, however, critics have argued that the benefits of post-hoc explanations are greatly exaggerated since they merely approximate, rather than replicate, the actual reasoning processes that black box systems take to arrive at their outputs. In this article, we aim to defend the value of post-hoc explanations against this recent critique. We argue that even if post-hoc explanations do not replicate the exact reasoning processes of black box systems, they can still improve users’ functional understanding of black box systems, increase the accuracy of clinician-AI teams, and assist clinicians in justifying their AI-informed decisions. While post-hoc explanations are not a “silver bullet” solution to the black box problem in medical AI, we conclude that they remain a useful strategy for addressing the black box problem in medical AI.

\bigskip

This is a pre-print of: Hatherley, Joshua, Lauritz Aastrup Munch, and Jens Christian Bjerring. 2025. In defence of post-hoc explanations in medical AI \textit{Hastings Center Report}, accepted March 1st 2025. \href{https://doi.org/10.1002/hast.4971}{10.1002/hast.4971}}

\begin{document}

\maketitle

\section{Introduction}

Medical AI has the potential to improve patient health and safety, reduce inefficiencies in medical decision-making, strengthen clinician-patient relationships, and increase the accessibility of high-quality medical care \citep{sparrow2019promise,topol2019high}. Increasingly, these systems are making their way into clinical practice. Indeed, the rate of regulatory approvals for AI systems designed to assist in all manner of clinical tasks – including diagnosis, risk-prediction, and surgical decision-making – have rapidly accelerated over the past 5-10 years \citep{zhu20222021}. While clinical adoption of medical AI systems remains nascent, clinical usage of these systems has increased exponentially since 2018 \citep{wu2024characterizing}. Recent developments in generative AI and foundation models only appear to be escalating these trends.

However, a longstanding concern with AI systems in high-stakes decision-making environments is that their inner workings are opaque \citep{brock2018learning,moor1985computer}. For deep neural networks and the like, not even the system’s designers themselves can scrutinise the reasoning behind each of these systems’ outputs. In medicine, this has prompted a range concerns relating to patient safety, trust, health equity, and patient autonomy, among other things \citep{bjerring2021artificial,hatherley2020limits,yoon2022machine}. In some cases, experts have even gone so far as to recommend that “black box” AI systems be banned from use in high-stakes decision-making environments entirely \citep{campolo2017ai}. 

Post-hoc explanation models refer to a suite of AI systems designed to reverse-engineer explanations for the outputs and operations of black box systems on the basis of their prior record of inputs and outputs. These explanations have been praised for their potential to address, or indeed, overcome the black box problem by rendering these systems “interpretable” or “explainable” \citep{ribeiro2016should}. But several critics have recently argued that post-hoc explanations are unlikely to benefit the users of black box medical AI systems since they do not replicate, but merely approximate, the actual reasoning processes of these systems \citep{babic2021beware,ghassemi2021false}. As a result, critics suggest that, not only do post-hoc explanations fail to improve users’ understanding of black box medical AI systems, they also fail to reduce the patient safety risks associated with their use, and in some cases, may even exacerbate these risks.

This article aims to defend the value of post-hoc explanations against these recent critiques. Despite their limitations, we argue that post-hoc explanations can still improve users’ functional understanding of black box systems, increase the accuracy of clinician-AI teams, and assist clinicians in justifying their AI-informed decisions, among other benefits. As a secondary aim, this article also aims to counteract what we refer to as the “silver bullet” thesis in Explainable AI (XAI) research. This thesis states that the black box problem is a single, unified problem that can be solved with a single solution. We argue that the black box problem is a multifaceted problem that demands varied strategies to address to the many risks it presents in medicine. While post-hoc explanations cannot “solve” the black box problem in medical AI, they are nevertheless a useful tool in mitigating it. Other strategies, both technical and non-technical, will be needed to overcome or sufficiently mitigate the black box problem. But the value of post-hoc explanations must not be overlooked for their inability to achieve these objectives in one fell swoop.

The structure of the remainder of the article is as follows. Section \ref{2} describes the black box problem in medical AI. Section \ref{3} gives an overview of XAI and post-hoc explanations. Section \ref{4} outlines the case against post-hoc explanations in medical AI. Section \ref{5} offers a defence of post-hoc explanations against these recent critiques.

\section{The black box problem}\label{2}

Many advanced AI systems – including deep neural networks, support vector machines and random forests – are black boxes. While stakeholders can scrutinise the inputs and outputs of these systems, their underlying reasoning process is largely resistant to human understanding. AI systems can be opaque for a variety of reasons, including intentional concealment, technical illiteracy, and model size or complexity \citep{burrell2016machine}. More specifically, however, AI systems can exhibit at least two different types of opacity: training opacity and inference opacity. Anders Søgaard claims that \textit{training opacity} occurs “when expert humans cannot, upon inspection, say how, in general, the parameters of the DNN were induced as a result of its training data” \citep[225]{sogaard2023opacity}. In contrast, \textit{inference opacity} occurs “when expert humans cannot, upon inspection, say why, in general, [a model] predicts an output y as a result of an input x” \citep[225]{sogaard2023opacity}.  

Machine learning systems often exhibit training opacity because learning algorithms are highly sensitive to tiny changes in the training dataset, including the order in which training data is presented to it. It is practically impossible, therefore, for users to predict the final state of the model from the data used to train it. On the other hand, inference opacity arises when a model's size – specifically, the number of parameters it includes – exceeds the scope of human short-term memory. For instance, examining a diagnostic AI system would not help a doctor understand the model's reasoning, as these models often contain millions of weighted parameters tied to specific textual or visual features that the model has learned during training.

This characteristic of AI systems gives rise to the so-called “black box problem” in medicine. According to Joseph Wadden, the black box problem “occurs whenever the reasons why an AI decision-maker has arrived at its decision are not currently understandable to the patient or those involved in the patient’s care because the system itself is not understandable to either of these agents” \citep[767]{wadden2022defining}. However, the black box problem has two distinct aspects: a technical aspect and an ethical aspect. 

The technical aspect of the black box problem concerns the computational challenges that must be overcome to “open the black box” of AI. \cite{guidotti2018survey} identify four such challenges.  First, the model explanation problem is that users cannot understand the underlying logic of a black box model as a whole. Unlike decision trees, for instance, support vector networks do not provide a chain of decision rules through which they proceed when classifying their outputs. Second, the outcome explanation problem is that users cannot understand the specific reasons underlying particular outputs. For example, black box systems do not provide explanations for why they classified a particular image of a skin lesion as malignant and another image as benign. Third, the model inspection problem is that users cannot understand specific properties of black box systems (e.g. the concepts or ‘features’ that influence their outputs). As Guidotti and coauthors observe, researchers aim to address this problem by “providing a representation (visual or textual) for understanding some specific property of the black box model or of its predictions” \citep[14]{guidotti2018survey}  Finally, the transparent box design problem differs from the other three problems; it concerns performance limitations in transparent models, rather than interpretability challenges in opaque systems. Transparent models can struggle to perform as accurately as black box systems. The aim here is to replicate the performance of a black box model using a transparent model (e.g. a decision-tree that approximate the rules an opaque model uses to classify inputs). 

In contrast, the ethical aspect of the black box problem refers to the specific ethical issues and risks associated with using black box systems in medical decision-making. \cite{krishnan2020against}, for example, identifies three such issues. According to the justification problem, if clinicians cannot understand the reasoning process behind an AI system’s classification or recommendation, they are not justified in accepting it \citep[see][]{shortliffe2018clinical}. According to the anti-discrimination problem, if clinicians cannot understand the reasoning process behind an AI system’s classification or recommendation, they cannot detect whether the output’s integrity has been compromised by algorithmic bias or spurious reasoning \citep[see][]{yoon2022machine}. Finally, according to the reconciliation problem, if clinicians cannot understand the reasoning process behind an AI system’s classification or recommendation, they cannot rationally resolve disagreements between themselves and the outputs of these systems \citep[see][]{grote2020ethics}.\footnote{Importantly, the list of ethical problems given in this paragraph is not exhaustive. For instance, concerns have also been raised about the impact of black box medical AI systems on patient autonomy, accountability for patient harm, and clinician-patient relationships \citep[see][]{smith2021clinical,sparrow2020high}.}

The ethical aspects of the black box problem can be addressed without solutions to the technical aspects of the problem. For example, providing users with information about the reliability and performance characteristics of a black box system can assist them in justifying their decisions to accept or override the outputs of these systems in certain decision-making scenarios \citep{duran2021afraid}. In the Explainable AI (XAI) literature, however, researchers tend to prioritise technical solutions, particularly post-hoc explanations. 

\section{What are post-hoc explanations?}\label{3}

The U.S. Defense Advance Research Projects Agency defines XAI as a research paradigm in AI that aims “to create a suite of new or modified ML techniques that produce explainable models that, when combined with effective explanation techniques, enable end users to understand, appropriately trust, and effectively manage the emerging generation of AI systems” \citep[45]{gunning2019xai}. The most popular approaches in XAI have thus far been post-hoc explanation models, which generate explanations by approximating the reasoning of a black box model on the basis of its prior record of inputs and outputs. 

Post-hoc explanation models come in a variety of different forms.\footnote{For a comprehensive review of post-hoc explanation models, see \cite{guidotti2018survey}.} For example, post-hoc explanation models can either be “model-specific” and “model-agnostic.” Model-specific post-hoc explanations can only be applied to particular classes of black box models (e.g. neural networks, support-vector machines, random forests, etc.). Saliency maps, for instance, are post-hoc explanation models that provide visual overlays highlighting the areas of an inputted image that were most influential in determining a system’s output \citep{jiang2013salient}. This type of post-hoc explanation is model-specific because it can only be used to explain neural networks and other gradient-based models. In contrast, model-agnostic approaches are those that can be applied to any type of black box model. Local interpretable model-agnostic explanations (LIME), for instance, are post-hoc explanation models that highlight visual or textual features of the input data that were most influential in determining its outputs \citep{ribeiro2016should}. This type of post-hoc explanation is model agnostic since it can be used to explain any class of black box model. 

Post-hoc explanations can also deliver global or local explanations. Global explanations attempt to explain the behaviour and operations of a black box model as a whole. For example, partial dependence plots are post-hoc explanations that provide a visualisation of the relationship between the input and output of a black box system in a reduced feature space \citep{greenwell2018simple}. Partial dependence plots provide global explanations because they provide insight into the relationship between specific features and predicted outcomes across the entire dataset, regardless of the values of other features. In contrast, local explanations attempt to explain the reasoning behind individual outputs. For example, SHapley Additive exPlanations (SHAP) offer a visualisation of the contribution of each feature to the difference between the actual prediction and the average baseline prediction for individual observations within the dataset \citep{nohara2022explanation}. SHAP provides local explanations because it computes the contribution of each feature to the difference between the actual prediction and the average baseline prediction for individual observations within the dataset. 

\section{The case against post-hoc explanations}\label{4}

Post-hoc explanations have been praised for their potential to improve users’ understanding and to promote appropriate reliance on black box systems, and to reduce the patient safety risks associated with their use \citep{combi2022manifesto}. Recently, however, several critics have argued that post-hoc explanations cannot make good on either of these promises \citep{babic2021beware, ghassemi2021false}.

The main reason for this is that post-hoc explanations are not “genuine” or “high-fidelity” XAI. Anantharaman Muralidharan and coauthors define high-fidelity XAI as systems in which “the representor generated by the secondary AI emulates, to a high degree, the way the primary actually arrived at its result” \citep[8]{muralidharan2024ai}. Yet as previously discussed, post-hoc explanations provide mere approximations of the actual reasoning processes taken by black box systems to arrive at their outputs. Any correspondence between a post-hoc explanation and the reasoning process of a black box system is merely a matter of chance. Indeed, Cynthia Rudin goes so far as to suggest that post-hoc explanations must be wrong: “If the explanation was completely faithful to what the original model computes, the explanation would equal the original model, and one would not need the original model in the first place, only the explanation” \citep[207]{rudin2019stop}.

As a result, critics argue that post-hoc explanations cannot improve users’ understanding of black box systems or their outputs. According to \cite{babic2021beware}, for instance, post-hoc explanations are a “fool’s gold” that provides mere “ersatz” understanding. For post-hoc explanation models do not provide genuine explanations, but mere rationalisations for the outputs of black box systems designed to appease users’ critical engagement with these systems. Such observations seem to be supported by a recent study in which post-hoc explanation models were found to generate intuitively reassuring explanations for even completely untrained machine learning models \citep{adebayo2018sanity}.  

Other critics argue that post-hoc explanations are often too shallow or uninformative to benefit users. According to Rudin, for instance, saliency maps do “not explain anything except where the network is looking” \citep[209]{rudin2019stop}. While these explanations can provide insight into what pixels of an image have influenced a system’s output, they cannot explain how these pixels influenced the models reasoning. Relatedly, other critics highlight that explanations are not the same as justifications. For example, a post-hoc explanation may explain how a model arrived at its output, but this does not demonstrate that the user ought to accept the output as true \citep{ghassemi2021false}.\footnote{Notably, this is a limitation of explanations more generally and is not specific to post-hoc explanations in particular.} Even if post-hoc explanations \textit{could} replicate the reasoning process a black box system has taken to arrive at an output, therefore, it would be insufficient to justify a user’s acceptance of this output.

Still other critics argue that post-hoc explanations are unlikely to promote appropriate reliance on black box systems. A range of empirical studies investigating the impact of post-hoc explanations on the performance of human-AI teams have produced concerning results. For example, \cite{kenny2021explaining} found that post-hoc explanations do not improve users’ trust in black box systems. \cite{eiband2019impact} found that even placebic explanations (i.e. explanations that offer no insight into a recommendation or request) can promote similar levels of self-reported trust on the outputs of black box systems compared to genuine explanations. \cite{adebayo2022post} also found that explanations may be unable to detect spurious correlations in the reasoning processes of black box systems.  

Finally, critics argue that post-hoc explanations may even exacerbate the very safety risks they are designed to address. Again, a variety of empirical studies investigating the impact of post-hoc explanations on the performance of human-AI teams have generated concerning findings. Automation bias, according to Mosier and coauthors, refers to “the tendency to use automated cues as a heuristic replacement for vigilant information seeking and processing” \citep[205]{mosier1996automation}. \cite{vered2023effects} found that, not only did post-hoc explanations fail to reduce automation bias in users, but in some cases actually increased it. Post-hoc explanations may also increase the difficulty users face when trying to detect algorithmic biases in black box medical AI systems, since post-hoc explanations have been found to exhibit biases themselves \citep{balagopalan2022road}. Finally, in some cases, providing post-hoc explanations for the outputs of black box systems has been found to result in worse overall human-AI team performance than simply eschewing these explanations altogether \citep{jesus2021can}.

Despite early optimism, a profound degree of pessimism concerning the potential for post-hoc explanations to address the black box problem now prevails, both in medicine and beyond \citep{swamy2023future}. In the next section, however, we defend the value of post-hoc explanations against these recent critiques.

\section{A defence of post-hoc explanations}\label{5}

Like human explanations, post-hoc explanations can benefit the users of black box systems in a variety of ways, even if they do not replicate the exact reasoning processes of these systems. After all, explanations from human experts often do not conform to the mechanistic reasoning process through which they arrived at their judgments, yet we confidently rely upon them regardless \citep{zerilli2019transparency}.

For example, post-hoc explanations can improve the performance of clinician-AI teams. In a recent study, \cite{senoner2024explainable} found that radiologist-AI teams were 4.7 percentage points more accurate in identifying lung lesions from chest x-ray images when saliency masks were used, compared to just the black box system alone. More research is needed to better understand the conditions under which post-hoc explanations improve clinician-AI performance. However, these findings demonstrate that, under some conditions, post-hoc explanations can have a significant positive impact on the accuracy of clinician-AI decision-making.

Relatedly, even if post-hoc explanations cannot reliably detect when a black box AI system relies on spurious correlations, as discussed previously, they can still improve users’ ability to discriminate between correct and incorrect outputs. In the aforementioned study, for instance, Senoner and coauthors found that saliency masks improved the performance of clinician-AI teams because “domain experts supported by explainable AI were more likely to follow AI predictions when they were accurate and more likely to overrule them when they were wrong” \citep[9]{senoner2024explainable}. Post-hoc explanations have also been found to reduce users’ overreliance on the outputs of AI systems in some instances \citep{vasconcelos2023explanations}.

As Páez points out, moreover, 

\begin{quote}
    models, idealizations, simulations, and thought experiments, […] play important roles in scientific understanding despite being literally false representations of their objects. In a similar vein, the methods and devices used to make black box models understandable need not be propositionally stated explanations and they need not be truthful representations of the models \citep[447]{paez2019pragmatic}. 
\end{quote}

In one influential study, for example, \cite{wang2021explanations} found that post-hoc explanations improved users’ ability to predict how the outputs of a model may change in response to changes in input data). The ability to predict a system’s outputs is commonly highlighted as a key component of explainability in AI \citep{chazette2021exploring,speith2024conceptualizing}. Doctors that can predict how changes in input data will affect a system’s outputs have some degree of causal understanding of the system. As such, they are better equipped to assess the weight these outputs ought to have in their decision-making, and to explain their AI-assisted reasoning process to patients. 

Furthermore, even if post-hoc explanations cannot justify a black box system’s outputs,\footnote{It is worth noting that good justifications need not correspond to the actual reasons for a decision. For example, suppose a manager rejects a proposed improved to the technical infrastructure of an organisation due to lack of budget or resources, when the actual reason is that they are concerned about how these changes would affect their role. Regardless of the manager’s actual reason, lack of budget or resources remains a good justification for rejecting the proposal. A problem, however, is that insincere justifications cannot be reliably action guiding, since the manager’s decision would not change if sufficient funds or resources became available \citep[see][]{babic2021beware}.} they can provide clinicians with evidence to justify their AI-informed decisions. As \cite{theunissen2022putting} have argued, post-hoc explanations can be combined with “institutional explanations” to make the former more meaningful to the users of these systems. These explanations include information about “the design decisions that went into the making of the [black box] system just as much as they are the technical specifications of the system: why should we trust the company and engineers designing it, what efforts have they been [sic] made to avoid bis, and how have they worked alongside end-users who will use these systems in their medical practice?” \citep[2]{theunissen2022putting}. The justificatory power of post-hoc explanations combined with institutional explanations can also be bolstered using non-XAI methods, e.g. by providing information about the reliability of the system, the data on which the system was trained, the system’s strengths and weaknesses, known “edge-cases,” and so on \citep[see][]{cai2019hello}.

While critics are also correct that post-hoc explanations can promote overreliance or exhibit algorithmic biases, these risks are not unique to post-hoc explanations. For human explanations, too, are often biased and promote overconfidence \citep{langer1978mindlessness, tetlock1998close}. Despite this, experts often rely on such explanations in high-stakes decision-making contexts such as medicine. It is unclear why critics think we ought to hold post-hoc explanations to a higher standard in this regard.

Such risks can also be mitigated. For example, biases in local explanations can be reduced through robust local training methods, such as the Just Train Twice method \citep{balagopalan2022road}. Misleading explanations can also be minimised through aggregating multiple explanations rather than providing users with a single explanation \citep{molnar2020general}. Finally, the risk of automation bias and overconfidence can be minimised through “cognitive forcing” techniques, which Buçinca and coauthors define as “interventions that are applied at the decision-making time to disrupt heuristic reasoning and thus cause the person to engage in analytical thinking” \citep[2]{buccinca2021trust}. For example, developers could force users either to wait 30 seconds or to make their own judgments before receiving an AI system’s recommendations and explanations. Developers could also incorporate the principles of “frictional AI” into the design of post-hoc explanation models, which Cabitza and coauthors define as “a general class of decision support aimed at achieving a useful compromise between the increase of decision effectiveness and the mitigation of cognitive risks, such as over-reliance, automation bias and deskilling” \citep[1]{cabitza2024never}. For example, developers could provide users with multiple AI-generated recommendations or explanations instead of a single output with the aim of forcing users to engage their critical faculties instead of deferring to the machine. 

Post-hoc explanations, therefore, need not replicate the reasoning processes that black box systems take to arrive at each of their outputs to benefit users. For despite this limitation, post-hoc explanations can still improve users’ functional understanding of black box systems, increase the accuracy of clinician-AI teams, and assist clinicians in justifying their AI-informed decisions. While post-hoc explanations can promote overconfidence or produce biased explanations, these risks are also characteristic of human explanations and can be mitigated through a variety of different strategies.

\bigskip

Post-hoc explanations are not a silver bullet solution to the black box problem in medical AI \citep{hatherley2024virtues}. This is to be expected. The black box problem is not a singular problem, but a collection of distinct technical challenges and ethical risks (see section \ref{2}). “Wicked” problems of this sort cannot be resolved in one fell swoop. Post-hoc explanations fall short of a complete solution, yet their utility in addressing the problem must not be overlooked. Addressing the black box problem requires multiple strategies, both technical and non-technical. Further research in the developing research area of mechanistic interpretability is promising \citep{bereska2024mechanistic}. But stronger focus on non-technical strategies is also needed. AI onboarding in clinical practice must be thorough, cautious, and reversible. Clinicians must be given regular opportunities to build and develop skills in AI-assisted decision-making. Medical AI systems must undergo detailed post-market surveillance of their performance and generalisability. The solution to the black box problem will not come from technological advancement alone; skilled users and effective oversight are vital.

\section*{Acknowledgments}

The research for this article was supported by a Carlsberg Foundation Young Researcher Fellowship (CF20-0257).

\bibliography{main}
\end{document}